\documentclass[11pt,a4paper]{article}

\usepackage{graphicx}
\usepackage{epsfig}

\usepackage{hyperref}
\usepackage{graphicx,epsf}
\usepackage{bm}
\usepackage{amsmath}

\begin{document}

\begin{center}
{\large{\textbf{Gyrokinetic investigation of Alfv\'en instabilities in the presence of turbulence}}}\\
\vspace{0.2 cm}
{\normalsize {A. Biancalani$^1$, A. Bottino$^1$, A. Di Siena$^1$, F. Jenko$^1$, P. Lauber$^1$, A. Mishchenko$^2$, I. Novikau$^1$,  and L. Villard$^3$.\\}}
\vspace{0.2 cm}
\small{$^1$ Max Planck Institute for Plasma Physics, 85748 Garching, Germany\\
$^2$ Max Planck Institute for Plasma Physics, 17491 Greifswald, Germany\\
$^3$ Ecole Polytechnique F\'ed\'erale de Lausanne, Swiss Plasma Center, CH-1015 Lausanne, Switzerland}\\
\vspace{0.1 cm}
{\footnotesize{contact of main author: \url{http://www2.ipp.mpg.de/~biancala/}}}
\end{center}
%
%
%
%
%
%

\date{\today}

\begin{abstract}
The global and electromagnetic gyrokinetic particle-in-cell code ORB5 is employed to investigate the self-consistent interactions between Alfv\'en modes (AM) and ion temperature gradient (ITG) turbulence in a magnetically confined plasma.  Here, an axisymmetric magnetic equilibrium with reversed shear and large aspect ratio is considered. An AM with toroidal mode number n=5 is driven unstable by introducing a population of suprathermal ions. Once the AM saturates in the presence of the fully developed turbulence, the ion heat flux is dominated by the AM and its main harmonics. ITG-induced transport is found to also be enhanced in the presence of the unstable AM.

\end{abstract}


\section{Introduction}

Magnetically confined plasmas are complex systems in which waves and instabilities at multiple spatial scales coexist and influence each other.
Important examples are microinstabilities, meso-scale zonal flows (ZF) and macroscopic MHD instabilities like Alfv\'en modes (AM), which mutually interact either due to direct coupling via wave-wave nonlinear interaction, or by indirect interaction mediated by suprathermal ion species, hereafter named ‘energetic particles’ (EP).
Microinstabilities, like ion-temperature-gradient (ITG) modes, are unstable due to the gradients of plasma temperature and density, and are particularly deleterious to the heat and particle confinement.
ZFs, i.e., ExB flows (primarily in the poloidal direction) associated with purely radial variations of the electrostatic potential, are usually observed in the presence of turbulence and can play the role of the dominant turbulence saturation mechanism~\cite{Hasegawa79}.
EPs are present in tokamak plasmas due to external heating mechanisms and to nuclear fusion reactions. AMs are eigenmodes of the shear Alfv\'en waves such as Global Alfv\'en Eigenmodes (GAE)~\cite{Appert82}, Toroidicity-induced Alfv\'en Eigenmodes (TAE)~\cite{Cheng85} or Beta-induced Alfv\'en Eigenmodes (BAE)~\cite{Chu92,Heibrink93}. These AMs can be driven unstable by the presence of EPs, which can then lead to a redistribution of the EP population~\cite{Chen16}. This can have consequences, inter alia, on plasma heating.

Due to the large computational cost of fully gyrokinetic (GK) simulations, the nonlinear dynamics of AMs has been studied in the past mostly with hybrid models treating the EPs and the thermal plasma (or part of it) with GK and fluid models, respectively. Recently, fully GK simulations have become affordable due to the availability of more powerful supercomputers, and more efficient numerical schemes (see for example Ref.~\cite{Cole17}).
By using a local (i.e., flux-tube) model, simulations of AMs in the presence of turbulence have also been performed, focusing on AMs in the limit of high toroidal mode number~\cite{Bass10}. In this Letter, 
we present, for the first time, global and fully GK simulations describing the self-consistent nonlinear interaction of AMs and ITG turbulence.

\section{Self-consistent global gyrokinetic simulations}

The stability of AMs at toroidal mode numbers $0\le n \le 9$ in an analytical equilibrium with reversed shear has recently been investigated~\cite{Biancalani19pop}.
In this Letter, we extend the previous study allowing higher-$n$ ITG modes to develop in the same equilibrium, and we study the self-consistent nonlinear interaction.
The gyrokinetic, global, electromagnetic, multi-species, particle-in-cell code ORB5 ~\cite{Jolliet07,Bottino11,Lanti19} used here, is based on a variational formulation of the electromagnetic gyrokinetic theory, which ensures appropriate conservation laws~\cite{Tronko19}. It uses state-of-the-art numerical schemes~\cite{Bottino15,Lanti19} that allow for transport time scales simulations.
The global character of ORB5, i.e., the resolution of the full radial extension of the global eigenmodes to scales comparable with the minor radius, makes ORB5 appropriate for studying low-$n$ AMs, without pushing towards the local limit of vanishing ratios of the ion Larmor radius to the tokamak minor radius.
ORB5 has been verified and benchmarked against the linear and nonlinear dynamics of AMs~\cite{Cole17,Koenies18,Taimourzadeh19}, ZFs~\cite{Biancalani14,Biancalani17pop}, and ITG modes~\cite{Goerler16,Tronko17}.
In this Letter, we discuss the interaction of ITG turbulence, AMs driven by EPs, and ZFs, as shown by ORB5 global self-consistent simulations.
All ion species are treated with a gyrokinetic model, whereas electrons are treated with a drift-kinetic model.
The pull-back scheme is used~\cite{Mishchenko18}.


\section{Equilibrium and identification of the instabilities}

The tokamak geometry, equilibrium magnetic field and plasma profiles are the same as in Ref.~\cite{Biancalani19pop}. The major radius is $R_0 = 10$ m, the minor radius is $a=1$ m, and the toroidal magnetic field at the axis is $B_0 = 3.0$ T. Circular concentric flux surfaces are considered. The reversed-shear safety factor has minumum value of $q=1.78$ at mid-radius.
The densities have a hyperbolic tangent profile, with gradient peaking at mid-radius.
The ion and electron temperatures are taken equal everywhere, $T_e=T_i$, with a hyperbolic tangent profile, with gradient peaking at mid-radius.
A value of $T_e$ at mid-radius corresponding to $\rho^* = \rho_s/a = 0.00571$, is chosen (with $\rho_s = \sqrt{T_e/m_i}/\Omega_i$ being the sound Larmor radius).
The electron thermal to magnetic pressure ratio at mid-radius is $\beta_e = 2\mu_0 P_e/B_0^2 = 5\cdot 10^{-4}$.
Ions are deuterons. 
The distribution function of the EP population is Maxwellian.  The EP population has $T_{EP}/T_e = 10$ at mid-radius, and has flat temperature profile. The EP concentration is $\langle n_{EP} \rangle/\langle n_e \rangle = 0.01$, where $\langle \, \rangle$ is the volume average (see Ref.~\cite{Biancalani19pop} for more details).

%

The analysis of this configuration for Alfv\'en stability has shown that a beta-induced Alfv\'en Eigenmode (BAE)~\cite{Chu92} with n=5, m=9 is the most unstable AM~\cite{Biancalani19pop}. The frequency and growth rate are $\omega_{BAE} = 2.4\cdot 10^{-3} \; \Omega_i$ and $\gamma_{BAE} = 0.68\cdot 10^{-3} \; \Omega_i$.
The localization is near the inner shear Alfv\'en wave continuum accumulation point, i.e. at $s=0.38$, with $s$ being the normalized poloidal flux radial coordinate.
The linear dynamics of micro-turbulence modes is dominated by ITG modes, with spectrum peaked around a toroidal mode number of n=26, where the frequency and growth rate are:
$\omega_{ITG} = 0.6\cdot 10^{-3}\, \Omega_i$, and $\gamma_{ITG} = 0.25\cdot 10^{-3} \, \Omega_i$. 
Note the separation of spatial and temporal scales between the AM dynamics and the ITG dynamics, which reflects the experimental observations in present tokamak devices. In particular, note that the ITG with n=5 has $\omega_{ITG,n=5} = 2.9\cdot 10^{-5} \, \Omega_i$ and $\gamma_{ITG,n=5} = 2.1\cdot 10^{-5} \, \Omega_i$, i.e., a linear frequency and growth rate two orders of magnitude smaller than the AM.


\section{AMs in the presence of turbulence}

Nonlinear simulations are performed with a filter allowing modes with toroidal mode numbers $0 \leq n \leq 40$ to evolve. These include the low-$n$ BAE, the whole spectrum of ITGs (mainly at high-$n$), and the ZF with $n$=0. No EPs are initialized at t=0, allowing the ITG modes to grow and establish a state of fully developed turbulence. After the turbulence saturation, the EPs are switched on, driving the BAE unstable.
In order to keep the number of species unchanged during the entire simulation, the EP species with  $\langle n_{EP} \rangle/\langle n_e \rangle = 0.01$ is initialized already at t=0 with $T_{EP}/T_e=1$ at mid-radius, and the EP temperature and density gradient are switched on only at the chosen time of $t=4.9\cdot 10^4 \, \Omega_i^{-1}$ (the density gradient only is checked not to sensibly modify the dynamics).
A spatial grid of (ns, nchi, nphi) = (256, 384, 192) points, a time step of dt=5 $\Omega_i^{-1}$, and a number of markers of $(N_i, N_e, N_{EP})$ = (2, 10, 2)$\cdot 10^8$ respectively for the thermal ions, electrons and EP are used for the turbulence simulations.
Dirichlet boundary conditions are applied to all components of the potentials at the axis and at the edge (except for the zonal potentials at the axis, for which Neumann boundary conditions are imposed).
A Krook (BGK) operator with $\gamma_{K}=1\cdot 10^{-4} \, \Omega_i$ is used to provide a restoring force for the equilibrium profiles, and for stabilizing the accumulation of turbulence energy.
A reduced mass ratio of $m_i/m_e=200$ is considered for simplicity (the saturation levels are found to be converged with this value).

\begin{figure}[b!]
\begin{center}
\includegraphics[width=0.52\textwidth]{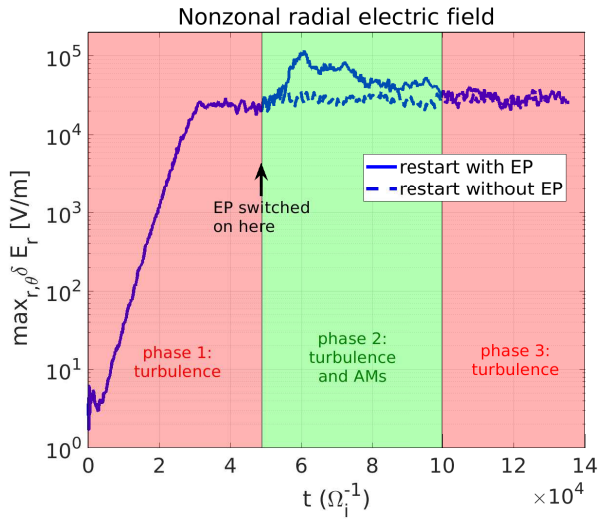}
\vskip -1em
\caption{\label{fig:time-evol} Evolution of the perturbed electric field. EPs are initialized after the turbulence has saturated, and an AM develops, saturates and relaxes back to the turbulent state (continuous line). For comparison, a simulation without EPs is shown as a dashed line, where only the ITG turbulence is present.}
\end{center} 
\end{figure}

\begin{figure}[t!]
\begin{center}
\includegraphics[width=0.46\textwidth]{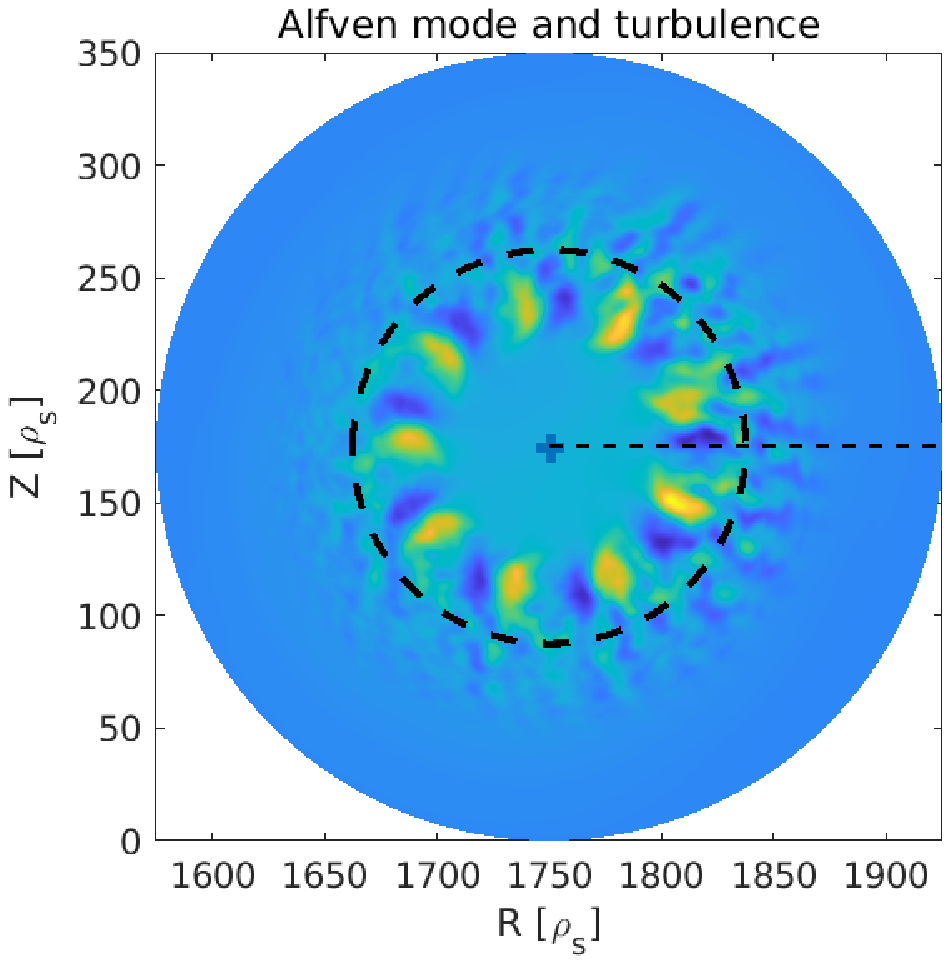}
\includegraphics[width=0.5\textwidth]{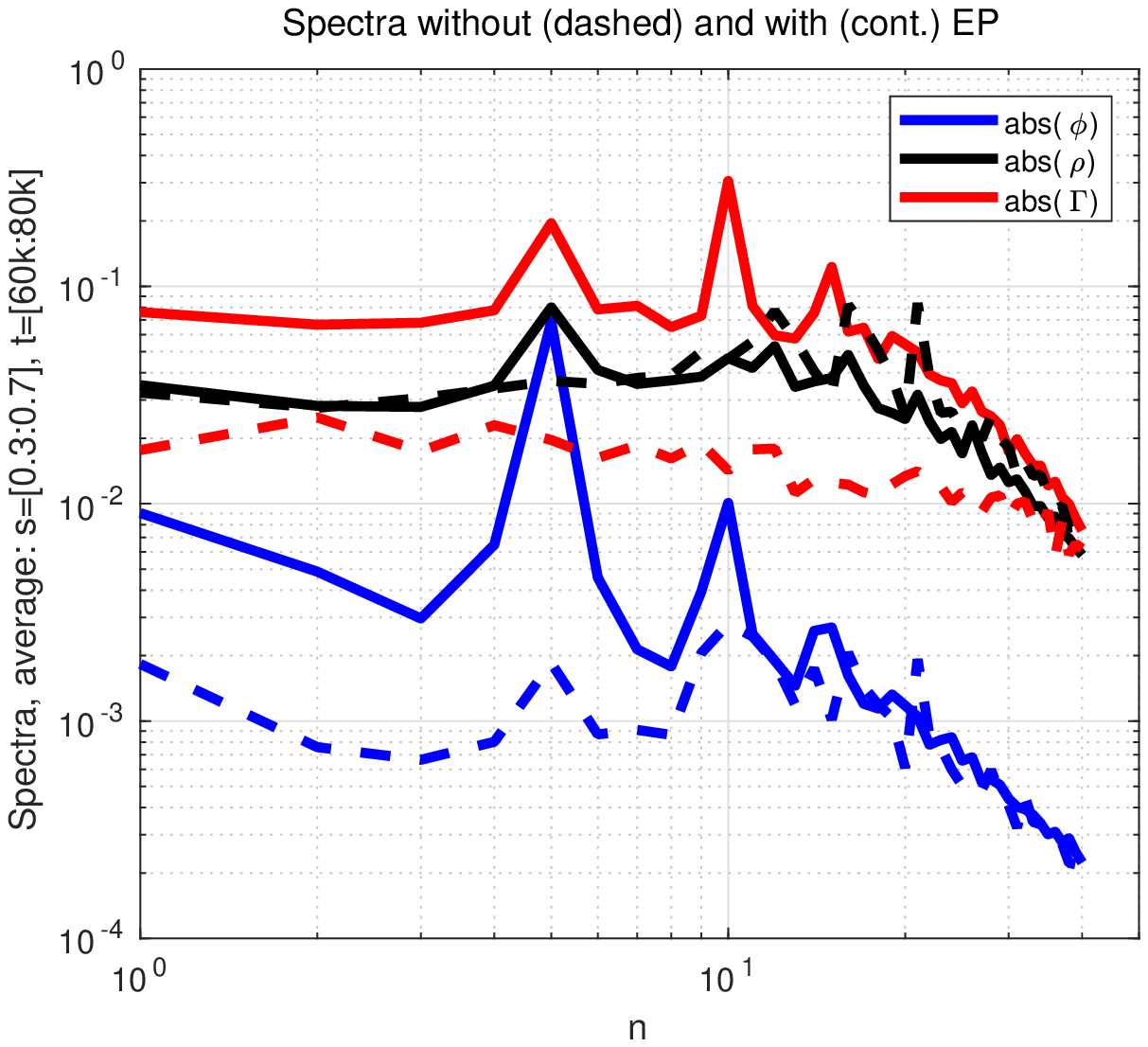}
\vskip -1em
\caption{\label{fig:ITG-BAE-struct} Poloidal cross section of the scalar potential (left), showing the BAE excited by an EP population in the presence of ITG turbulence. Toroidal mode number spectra (right) of scalar potential $\phi$, density $\rho$, and ion heat flux $\Gamma$.}
\end{center} 
\end{figure}

The evolution in time of the perturbations can be investigated by measuring the radial electric field. In particular, the nonzonal (i.e., oscillating in the toroidal and poloidal angles) radial electric field gives the amplitude of the modes with helicity, like ITGs and AMs. This is shown in Fig.~\ref{fig:time-evol}. For completeness, a snapshot of the scalar potential in the poloidal plane, showing the coexistence of the BAE and turbulence, is shown in Fig.~\ref{fig:ITG-BAE-struct}. The ITG microinstabilities develop due to the equilibrium ion temperature gradient, and form a turbulent state (phase 1).
The EPs are switched on at $t=4.9\cdot 10^4 \Omega_i^{-1}$. The EP density gradient drives a BAE unstable (phase 2).
The EP density radial profile is modified by the AM during the nonlinear phase. In particular, the absolute value of the EP density gradient decreases at the resonant radial location. This reduces the drive and yields the BAE saturation. 
After the saturation, the EPs are further redistributed, and the BAE amplitude slowly decreases. Eventually, the BAE vanishes, and the perturbed radial electric field decreases back to levels given by the ITG turbulence (phase 3). Zonal flows are also present in the simulations, being excited by turbulence via modulational instability and by the AMs via forced-driven excitation. The zonal flow level is found to be one order of magnitude higher in the case with EPs, with respect to the case without EPs, due to the effective forced-driven excitation.
Simulations with a different EP concentration, density gradient localization, or temperature, are observed to drive AMs at different amplitude, localization, frequency and mode number.


\section{Heat transport}

Both ITG turbulence and AMs are known to induce cross-field heat transport. In the simulations presented here, both of these effects are co-existing, raising two general questions: What is the relative importance of each of these two mechanisms, and do they interfere in any way?

The toroidal mode number spectrum of the ion heat flux (for the details of the diagnostic, see Ref.~\cite{Hayward19})
for a simulation where the EPs are switched on can be seen as a the continuous red line in Fig.~\ref{fig:ITG-BAE-struct}. For comparison, the same spectrum for a simulation without EPs is shown as a dashed red line. A sensible general trend that we find is that the heat fluxes are higher in the simulation with EPs. This is explained by the fact that the EP density profile consitutes an additional source of free energy to the system.
%
%

\begin{figure}[t!]
\begin{center}
\includegraphics[width=0.46\textwidth]{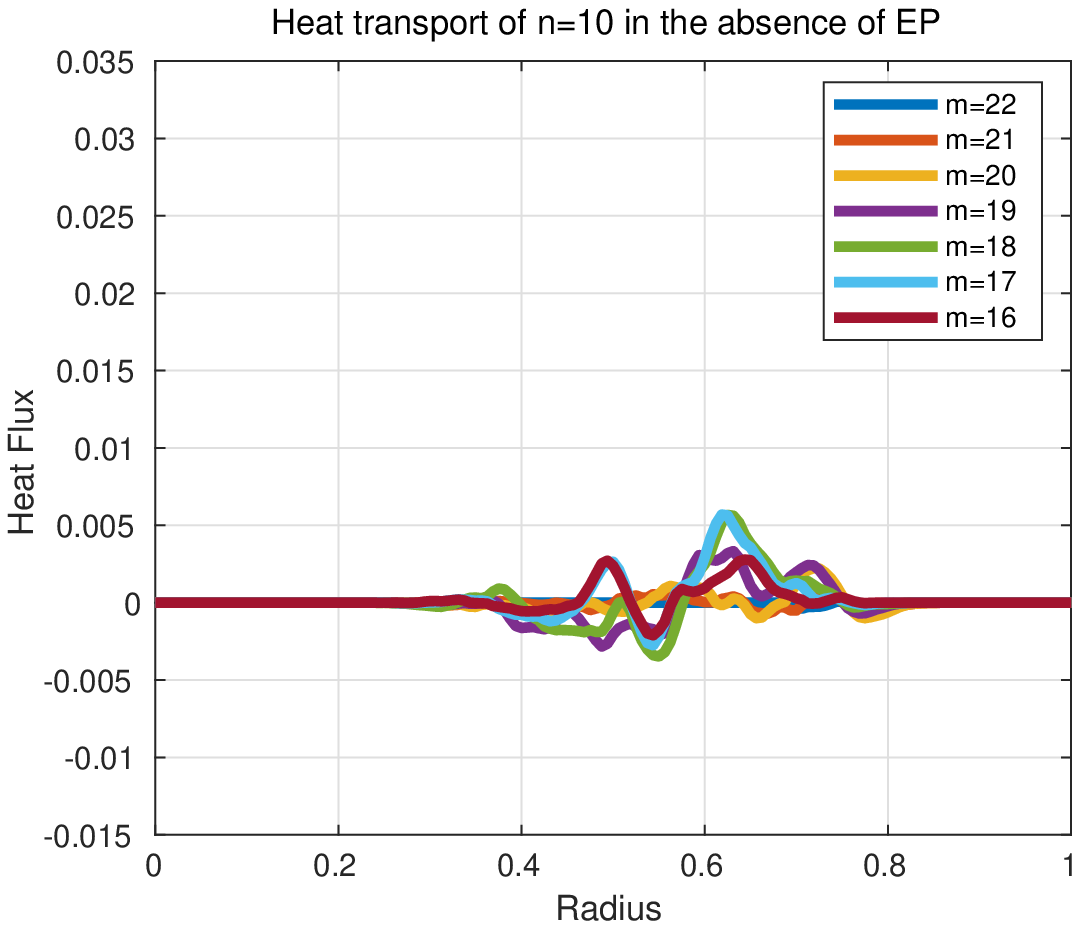}
\includegraphics[width=0.46\textwidth]{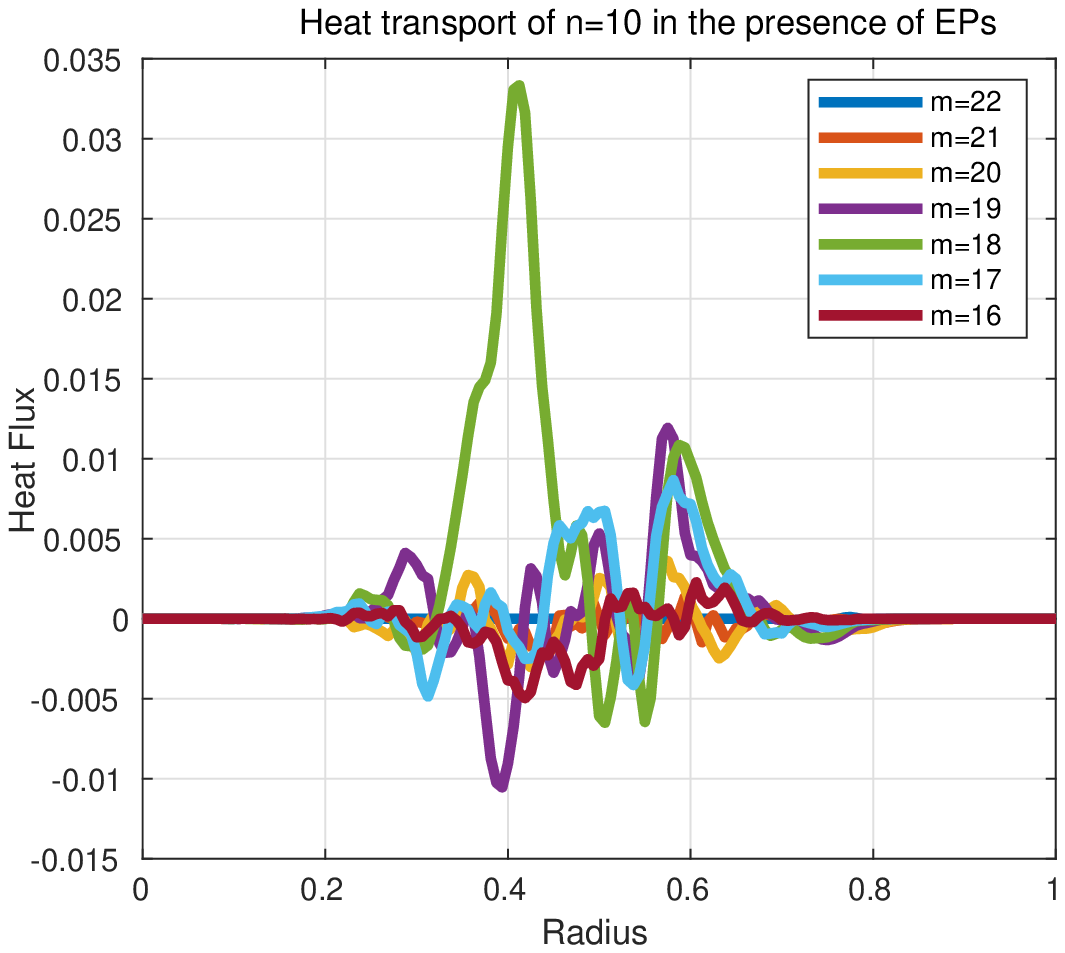}
\caption{\label{fig:n10woEP}
Radial structures of heat flux for $n=10$, and for different poloidal modes, for a simulation without EPs (left) and with EPs (right). Without EPs, the polarization with equal contribution of the different poloidal components identifies the ITG dynamics. With EPs, the polarization with dominant $m=18$ identifies the BAE second harmonics, on top of the ITG turbulence.
}
\end{center} 
\end{figure}

We observe that the spectrum is dominated by the modes $n=5$, $n=10$ and $n=15$, namely the first, second and third harmonics of the main BAE mode.
The contributions of the different poloidal mode numbers to the heat fluxes for the dominant mode, namely the mode with $n=10$, is depicted for a simulation without and with EPs, respectively in Fig.~\ref{fig:n10woEP} and Fig.~\ref{fig:n10woEP}.
In the case of the simulation without EPs, a broad spectrum of poloidal components shows the characteristic polarization of the ITG mode.
On the other hand, for the case of the simulation with EPs, the heat flux is dominated by the poloidal mode $m=18$.  Note also that the peak of the $m=18$ component is at the position of the peak of the BAE field. These two are signatures of the second harmonics of the BAE. Therefore, we can state that the ITG dynamics is subdominant in the simulation with EPs, with respect to the BAE, in carrying the heat transport.
The reason why BAEs can have such strong interaction with the thermal ion heat fluxes is identified in their relatively low frequency, with respect to AMs like toroidicity induced Alfv\'en eigenmodes. This relatively low frequency increases the important of resonances with thermal ions, which is less important for TAEs.
It is also important to note that the other poloidal components (i.e., $m\ne18$) are increased in the simulation with EPs, in comparison with the simulation without EPs. This means that the BAE second harmonic is efficient in modifying the dynamics of the ITG of $n=10$, due to the nonlinear interaction. This is an example of cross-scale interaction, with the AM being the macro-scale mode, and the ITG turbulence being constituted mainly by micro-instabilities.

It is worthy to note that not only the low-$n$ part of the heat flux spectrum, i.e., the BAEs, has higher levels in the presence of EPs, but also the higher-$n$ part, i.e., the ITGs. This is of interest for the question of how the presence of EPs modifies the ITG turbulence. In this regime, we can state that an EP population driving a BAE linearly unstable, affects the turbulence dynamics by increasing the heat transport.
The fluctuation amplitude of the scalar potential $\phi$ and density $\rho$ for the different toroidal mode numbers is also shown in Fig.~\ref{fig:ITG-BAE-struct} (respectively with blue lines and blue black lines). Differently from the heat flux, the fluctuation amplitude given by the scalar potential is shown not to be sensibly modified by the presence of EPs in the range of toroidal mode numbers of the ITG. Moreover, the fluctuation amplitude given by the perturbed density is found to be decreased by the presence of EPs.

\section{Conclusions and discussion}

In this Letter, we have investigated the nonlinear dynamics of an unstable BAE in the presence of fully developed ITG turbulence by means of global self-consistent gyrokinetic simulations. The nonlinear saturation of the n=5 BAE can be attributed to a radial redistribution of the EP population, causing a decrease of the drive. The BAE saturation level was found to be consistent with the one obtained in the absence of turbulence. The BAE and its first harmonics, namely the modes with $n = 10$ and $n = 15$, were found to dominate the ion heat flux. Such a strong contribution of the BAE to the thermal transport of the plasma, which was underestimated in the past, is explained by noting that the BAE has a low frequency with respect to other kinds of AMs, and is therefore capable of resonating with characteristic frequencies of the thermal ions. The heat flux caused by ITG modes around $n = 10$ also increases, indicating that there are nonlinear interactions between the macro-scale BAE and the micro-scale ITG modes. For comparison, we have performed the same simulations by filtering out the ZFs, and we have observed a sensibly different saturation level of the BAE (in this regime, smaller, consistently with Ref.~\cite{Biancalani19pop}) and a much smaller effect of the EP-driven BAE on the ITGs.

The results presented here illustrate how modes at different spatial scales can interact nonlinearly in a system as complex as a tokamak plasma, and therefore how a multiscale investigation is mandatory for quantitative predictions. In particular, a macro-scale BAE, meso-scale ZFs, and micro-instabilities like ITG modes, can coexist and contribute in different ways to the ion heat transport. Having demonstrated the feasibility of global self-consistent gyrokinetic simulations in this context, comprehensive theoretical studies of burning plasmas can be envisioned, with the goal to develop a predictive capability for ITER and future fusion power plants.

\section*{Acknowledgments}
Interesting discussions with F. Zonca, Z. Qiu, E. Poli, A. Zocco, T. G\"orler, T. Hayward-Schneider, \"O. G\"urcan, P. Morel, E. Lanti, N. Ohana, and F. Vannini are gratefully acknowledged. This work has been carried out within the framework of the EUROfusion Consortium and has received funding from the Euratom research and training program 2014-2018 and 2019-2020 under grant agreement No 633053 within the framework of the {\emph{Multiscale Energetic particle Transport}} (MET) European Eurofusion Project. The views and opinions expressed herein do not necessarily reflect those of the European Commission. Simulations were performed on the HPC-Marconi supercomputer within the framework of the ORBFAST and OrbZONE projects.

\end{document}